\documentclass[aps,prc,twocolumn,floatfix,showpacs,a4paper,nofootinbib,amsmath,amssymb]{revtex4-1}
\usepackage{graphicx}
\usepackage{color}
\usepackage{dcolumn}
\usepackage{bm}
\usepackage{amsmath}
\usepackage[section]{placeins}
\newcommand{\be}{\begin{equation}}
\newcommand{\ee}{\end{equation}}
\newcommand{\ba}{\begin{eqnarray}}
\newcommand{\ea}{\end{eqnarray}}
\newcommand{\bd}{\begin{displaymath}}
\newcommand{\ed}{\end{displaymath}}
\newcommand{\nnb}{\nonumber \\}

\begin{document}

\title{Radiation dominated implosion}
\author{
Susanne F. Spinnangr$^{1,2}$, 
Istv\'an Papp$^{2,3}$ and L\'aszl\'o P. Csernai$^{1,2}$}
\affiliation{
$^1$Dept. of Physics and Technology, Univ. of Bergen, Norway\\
$^2$Sustainability Center, Institute of Advanced Studies, K{\H o}szeg, Hungary\\
$^3$Dept. of Physics, Babes-Bolyai University, Cluj, Romania}

\begin{abstract}
Inertical Confinement Fusion configuration model is analized for direct 
ignition without an ablator.
The compression of the target pellet is neglected and rapid volume ignition is 
achieved by a laser pulse, which is as short as the penetraton time of the light
across the pellet. The reflectivity of the target is assumed to be negligible, and
the absorptivity is constant so that  the light pulse can reach the opposite side
of the pellet. The necessary pulse length and pulse strength is calculated.
\end{abstract}

\date{\today}


\maketitle

\section{Introduction}

We consider a spherical pellet of Deuterium-Tritium (DT) fuel
of initial outer radius of 1143 $\mu$m. At the National Ignition 
Facility (NIF) the pellet had a hole in the middle, to reach
better compression, and it had a thin "ablator" layer, which reflected
the incoming light
\cite{Lindl1998,Lindl2004,Haan2011}. 
In this experiment the target
capsule was indirectly ignited by the thermal radiation  coming from the gold
Hohlaum. The Hohlraum was heated by the radiation of 192 lasers
 \cite{RBOAH16,Nora2015}.
In the direct drive OMEGA laser experiment
the initial radius of the DT capsule was smaller, $R= 430\ \mu$m
\cite{RBOAH16}.
The incoming and reflected light exercised a pressure
and compressed the pellet at NIF, to about $R= 80\ \mu$m 
just before ignition. At this moment the hole in the capsule is already
filled in and the target density was compressed to about 300-700 g/cm$^3$
\cite{NIF11,RHL16}. Then this target showed the development of Rayleigh-Taylor
instabilities, which reduced the efficiency of ignition.

The initial compression pulse ("low foot")  had lower frequency of or
longer wavelength of 100-300 nm, which
therefore had a higher reflectivity on the target, and led to compression.
The reflectivity of light is high ( $> 0.6$ ) for lower frequency light and
decreasing with increasing frequencies. It becomes negligible at 
$\hbar \omega = 1 $ keV. The compression pulse
was followed by a shorter, higher frequency ignition pulse.
The higher frequency pulse has negligible reflectivity and
decreasing absorptivity, having  
$\alpha_K = 10^6$ cm$^{-1}$ at $\hbar \omega = 20 $ eV and
$\alpha_K = 10$ cm$^{-1}$ at $\hbar \omega = 1 $ keV.
If we take a higher frequency, shorter wavelength
of 20 nm in the X-ray range then the absorptivity of the DT fuel is
about $\alpha_K = 10^4$ cm$^{-1}$  \cite{HuCoGo14}. 
This means that the full pulse energy 
is absorbed in $10^{-4}$ cm $= 1 \mu$m. That is in a thin surface layer.
The internal domain is heated up due to adiabatic compression, up
to ignition, but the major part of the approximately 10 $\mu$m thin 
compressed surface layer remains cold and only a 1 $\mu$m is heated 
up at the outside surface. See Fig. 8 of Ref.~\cite{HuCoGo14}.

\section{Considerations for the target}

In our following consideration we could take (a) a compressed smaller
initial state of radius $R= 80\ \mu$m, which is then not compressed further but 
heated up further with a short penetrating light pulse. Alternatively
we could (b) consider a solid, not compressed ball of the same amount 
of DT fuel, which
is then made transparent and ignited by a laser pulse without
compression of radius $R= 640\ \mu$m,.  In this second case due to the  
smaller density we will need a more energetic short pulse but also
8 times longer.

Reference \cite{LCJ16} used a similar size target, with an outer ablator 
layer and initial compression. However, they used a special 
cone-in-shell configuration of the target. Through doping the target 
with Cu they were able to image the K-shell radiation of the target 
when it was radiated by an ultraviolet driver beam. From their images 
in Figure 2 d-f  of \cite{LCJ16}, we see that  ignition is achieved in an area of 
approximately 50 $\mu$m  radius from the center of the target. 
By using a high-contrast laser and a 40 $\mu$m cone tip they 
were able to increase the fast electron coupling to the core from
$<5\%$ to 10-15\% by an increase in the core-density and decrease in 
the source-to-core distance. If this laser-to-electron conversion 
efficiency would be further increased, the total laser energy coupled 
to the core would also increase above 15\%, which may be good if we 
want to achieve fast-ignition inertial-confinement-fusion. 

Here we can consider another configuration, without ablator layer
and without pre-compression, using the early examples in Refs.
\cite{C87,CS14}.
Then we have a more dilute target
fuel with about 640 $\mu$m radius, case (b). 
With a deuterium-tritium ice as fuel, 
the target density is 1.062 g/cm$^3$. This target a priory
has smaller absorptivity. If we want to absorb the whole energy
of the incoming laser light on $\sim$ 1.3 mm length, we need an
absorptivity of  $\alpha_K \approx 8 $ cm$^{-1}$. This is about the 
absorptivity of DT fuel for soft X-ray radiation of 1 nm wavelength.
Longer wavelength radiation would have a larger absorptivity, and
would be absorbed in the outside layers of the pellet.

\section{Simplified model and its evaluation}

Consider a spherical piece of matter (E), which is sufficiently
transparent for radiation.  The absorptivity of the  target matter is 
considered to be constant, such that the total energy of
the incoming light is observed fully when the light reaches the
opposite edge of the spherical target. 
 This matter undergoes an exothermic
reaction if its temperature exceeds $T_c$.

\begin{figure}[h]  
\begin{center}
\resizebox{0.8\columnwidth}{!}
{\includegraphics{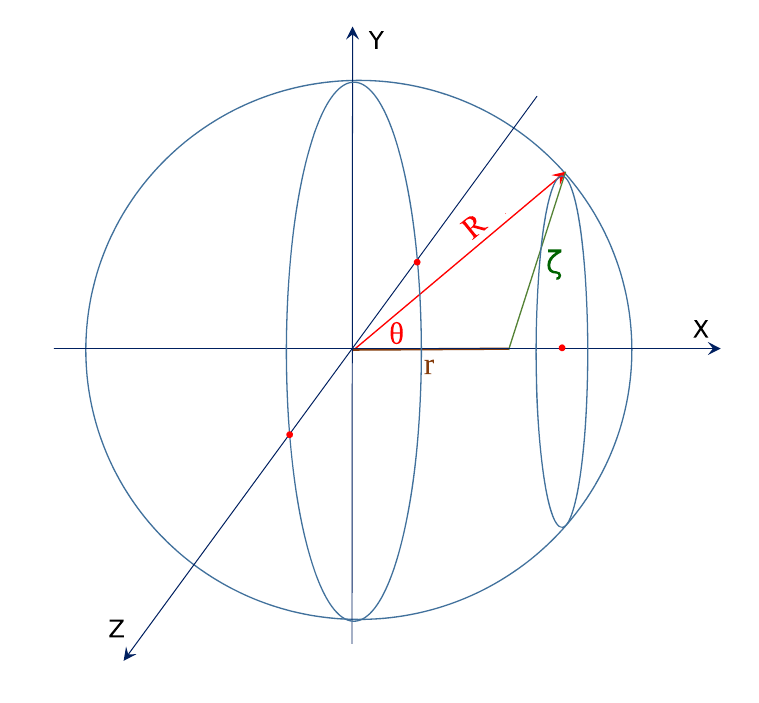}}
\caption{
(color online) 
The sphere of the fuel, with an internal point at radius $r$.
Let us chose the $x$-axis so that it passes through the point at
$r$ and the center of the sphere. Then let us chose a point on the
sphere, and the angle of this point from the $x$-axis is denoted 
by $\Theta$. Then the length between this surface point and the
internal point at $r$ is $\zeta = ( R^2 + r^2  -2 R r \cos \Theta)^{1/2}$. 
The propagation time from the
surface point to the point at $r$ equals $\tau = \zeta/$c.
}
\label{F-1}
\end{center}
\end{figure}

The target matter is surrounded by an set of spherically distributed 
lasers, 
which emit the radiation necessary to heat up E. We are neglecting
the expansion of the outer shell inwards as well as the 
expansion of the core, so that the core radius $R$ is constant.
We will measure the length in units of $\mu$m, and the time in units of 
$\mu$m/c.

We intend to calculate the temperature, $T$, distribution within the 
sphere, as a function of time, $t$, and the radial distance from
the center of the sphere, i.e. radius $r$.  We have two steps of the
evaluation: \\
(i) In the 1st step we calculate that from the outside
surface of the sphere how, much energy can reach a given point at
$r$. Here we have to take into account that the outside thermal
radiation starts at time, $t=0$, so there is no radiation before, and 
only those parts of the outside surface can reach a point inside the
sphere at time $t$, and which are on the backward light-cone of the
point at $r$ and time $t$. The integral for the energy density
reaching the point from this part of the two 
dimensional outside surface of the sphere in unit time 
interval, $dt$, is $dU(r,t)/dt$.\\
(ii) Then we have to add up the accumulated radiation at position
$r$, for the previously obtained energy and to obtain the 
time dependence of the temperature distribution, $T(r,t)$, we
have to integrate  $dU(r,t)/dt$ from  $t=0$, for each spatial position.\\
We perform the surface integral of step (i) in terms of integration
for the proper time of the radiation with a delta function, selecting
the surface element, which can reach the given internal point at a time.

Let us study a point within the sphere, at a distance $r$ from
the center. Choose the $x$-axis passing through this point and the
center of the sphere. See Fig. \ref{F-1}.

The surface area of a ring of the sphere at the selected 
polar angle $\Theta$ is $dS = 2 \pi R^2 \sin \Theta \, d\theta$.

Step (i):\\
At a point at $r$ we receive radiation from a layer edge ribbon at 
time $\tau$. The radiation
at distance $\zeta$ is decreasing as $1/\zeta^2$. The total radiation 
reaching point $r$ from the ribbon at $\Theta$ is 
\be
dU(r,t) \propto  \frac{1}{\zeta^2} \delta(\zeta{-}\sqrt{R^2{+}r^2{-}2rR\cos\Theta})\ ,
\ee
where $\tau=\zeta/$c, 
and we should integrate this for the surface of all ribbons.

The average intensity of thermal radiation reaching the surface of 
the pellet amounts to 
$Q$ per unit surface ($\mu$m$^2$) and unit time ($\mu$m/c).
Let us take a typical value for the energy of the total ignition
pulse to be  2\, MJ, in time 10\, ps,  then  \
$Q= 2 {\rm MJ}\ (4 \pi \cdot 640 \mu {\rm m})^{-2}\ (10 {\rm ps})^{-1}$\  or\ 
$ Q= 3.08  \cdot 10^{19}\ {\rm W / cm}^{2} =
1.03 \cdot 10^{9}\    {\rm J\,c / cm}^{3} $.

\begin{figure}[h]  
\begin{center}
\resizebox{0.8\columnwidth}{!}
{\includegraphics{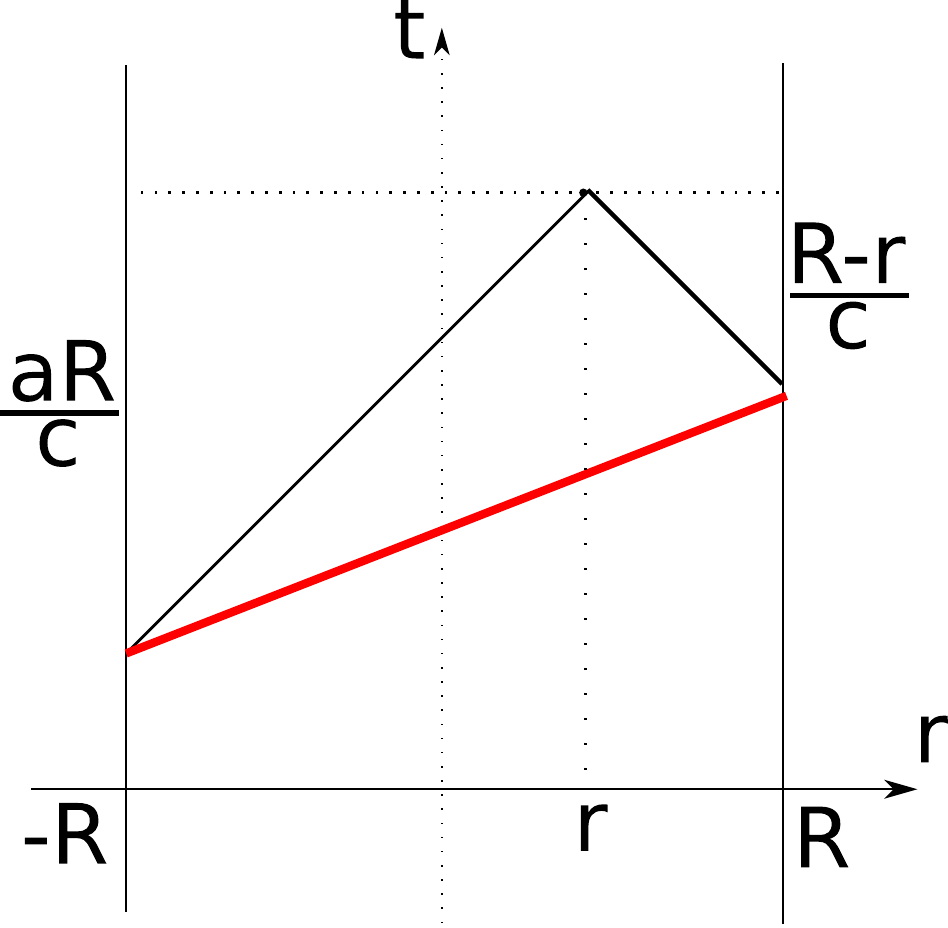}}
\caption{
(color online) 
The boundaries of the integration domains depending on
$r$ and $t$. The domain for the smallest $\tau$ -values
cannot receive radiation, Eq. (2c), because the radiation started at
$(R-r)/$c earlier and it reaches the internal point at $r$ later, 
At the same time the radiation from the opposite 
side reaches the point $r$ also in time $aR/$c. The contour of the
intersection of the backward light-cone with the surface of the 
sphere is indicated with the thick, red line.
When the 
time from the momentum of ignition is longer than
$t=2R$/c the radiation reaches the matter from all
sides at every location $r$. At earlier times the upper boundary
of integration should be evaluated. See Eq. (\ref{E4}).
}
\label{F-2}
\end{center}
\end{figure}

Up to a given time $t$, the light can reach a space-time 
point ($r,t$), inside the 
sphere from different points  of the outside surface, which were emitted in
different times. At early times it may be that none of the surface points are
within the backward light-cone of the point ($r,t$). At later times, 
from part of the
surface points the light can reach ($r,t$),  while at times larger 
than $2R/$c
all internal points can be reached from any surface point of the sphere. Thus,
we calculate first what energy density, $U(r,t)$, we get at a space-time 
point ($r,t$), from earlier
times.
At a 
given point at $r$ measured from the center of the sphere
(assuming that a constant fraction, $\alpha_K$, of the radiation energy is 
absorbed in unit length):
\footnote{
We are using the relation
$\delta[g(x)] = \sum_i [ (1/|g'(a_i)|) \delta(x-a_i)$ where $a_i$-s are the 
roots of $g(x)= \zeta - \sqrt{R^2+r^2-2rRx}$, i.e. $g(a_1)=0$. Now 
$a_1 = (R^2+r^2-\zeta^2)/(2rR)$  and
$g'(x) = rR/\zeta$ so that the integrand is $\zeta/(\zeta^2 rR) = 1/ (rR \zeta)$.
The variable $\zeta$ depends on $x$ (or $\Theta$), so we should set the 
integral boundaries in terms of $\zeta$ accordingly.
} 

\ba
&& dU(r,t) = 
\nnb
&&
\alpha_K Q\! \int_0^t \!\!\!\! d\tau\, 
2\pi R^3 \int_0^{\pi} \!\!\!\!d \cos\Theta\,
\frac{\delta(\zeta{-}\sqrt{R^2{+}r^2{-}2rR\cos\Theta})}{R^2+r^2-2rR\cos\Theta}
\nnb
&& = \alpha_K Q\! \int_0^t \!\!\! d\tau\, 
2\pi R^3 \int_{1}^{-1} \!\!\!d x\,
\frac{\delta(\zeta{-}\sqrt{R^2{+}r^2{-}2rRx})}{R^2+r^2-2rRx}
\nnb
&& = 2\pi R^3 \alpha_K Q \cdot (Rr)^{-1} 
\int_{(R{-}r)/c}^{aR/c} \frac{d\tau}{\tau{\rm c}} \ ,
\ea
where the integral over $dx$ gives $1/(Rr\zeta)=(R\,r\,\tau\,c)^{-1}$. 
The time, $d\tau$, integral runs from the nearest point of the backward 
light cone to the surface of the sphere to the furthest point, $aR/c$. 
Here the parameter $a$ will be described later.   See Fig. \ref{F-2}.

Now we introduce a new, dimensionless time variable:
$$
q \equiv \tau {\rm c}/R 
$$.
Thus,
\ba
dU(r,t) &=& 
2\pi R^3 \alpha_K Q \cdot (rR{\rm c})^{-1} \int_{1{-}r/R}^{a} \frac{dq}{q}
\nnb
&=&
 2\pi R^2 \alpha_K Q \cdot (r{\rm c})^{-1}\, [ \ln (q) ]_{1{-}r/R}^{a}
\ea
%
where 
$\zeta= \tau$c and
$a$ is the upper boundary of the integral over  the dimensionless time $dq$
\be
a = \left\{
\begin{array}{ll}
	1-r/R\ , & q<1{-}r/R \\
	q   \ , & 1{-}r/R<q<1{+}r/R \\
	1+r/R\ , & q>1{+}r/R
\end{array}
\right.
\label{E4}
\ee

Here actually the integral over $dq$ is adding up the contributions
of those surface elements of the sphere, from where radiation reaches the
internal point at $r$ at the same dimensionless time $q$. In the first case the radiation
does not reach the point at $r$ then, in the second part the radiation 
from the closest point of the sphere reaches $r$ but from the opposite point
not yet, in the third case radiation reaches $r$ from all sides.

Thus the energy deposited in unit time at dimensionless time $q$ is
\ba
&&dU(r,q) = \frac{2 \pi R^2 \alpha_K Q}{rc} \times
\nnb
 &&\left\{
\begin{array}{ll}
	\ln[(1+r/R)/(1-r/R)] , & q>1{+}r/R \\
	\ln[q/(1-r/R)] , & 1{-}r/R<q<1{+}r/R \\
	0 , & q<1{-}r/R
\end{array}
\right.
\nnb
\ea

Step (ii):\\
Neglecting the compression and assuming constant specific heat $c_v$,
we get that $k_B\,dT = \frac{1}{n\,c_v} dU\cdot dq$, where $k_B$ is the 
Boltzmann constant, and so
\ba
&&k_B\, T(r,t) =  \frac{1}{n\,c_v} \int_0^{t{\rm c}/R} dq \cdot dU(r,q) = 
\frac{2 \pi R^2 \alpha_K Q}{n\,c_v\, r c} \times
\nnb
&&\left\{  
\begin{array}{ll}
\left[q\,\ln\left(1+r/R)/(1-r/R)\right)\right]_{1+r/R}^{tc/R }+ \\
\ \ \ \	(1+r/R)\,\ln[(1+r/R)/(1-r/R)] - 2r/R\, ,\\
\ \ \ \ \ \ \ \ \ \ \ {\rm if:}\ \ \ \ \  tc/R>1{+}r/R\\
\left[q\,\ln\left(q/(1-r/R)\right) -q\right]_{1-r/R}^{tc/R}\, ,  \\
  \ \ \ \ \ \ \ \ \ \ {\rm if:}\ \ \ \ \  1{-}r/R<tc/R<1{+}r/R \\
	0\, , \ \ \ \ \ \ \ {\rm if:}\ \ \ \ \ \ tc/R<1{-}r/R
\end{array}
\right.
\nnb
&& = H \cdot \frac{R^2}{r} \times
\nnb
&&\left\{  
\begin{array}{ll}
	tc/R\,\ln[(1+r/R)/(1-r/R)] -2r/R\, , \\
\ \ \ \ \ \ \ \ \ \ \ {\rm if:}\ \ \ \ \  \ tc/R>1{+}r/R \\
	tc/R\,\ln[tc/R/(1-r/R)] -tc/R+1-r/R\, , \\
\ \ \ \ \ \ \ \ \ \ \ {\rm if:}\ \ \ \ \  1{-}r/R<tc/R<1{+}r/R \\
	0\, , \ \ \ \ \ \ \ \ {\rm if:}\ \ \ \ \  \ tc/R<1{-}r/R ,
\end{array}
\right.
\nnb
\ea
where the number density of uncompressed DT ice is \\
$n =  3.045 \cdot 10^{22}$ cm$^{-3}$, and the leading constant, $H$, is
\be
H \equiv \frac{2 \pi  Q}{c \, c_V}\, \frac{\alpha_K}{n} 
= 6.8 \cdot 10^{-13} {\rm J / cm}.
\ee

Thus it follows:
\be
k_B\,T(r,t) \propto \left\{
\begin{array}{ll}
0 \ ,\ \ \ \ \ {\rm if:}\ \ \ tc/R{<}1{-}r/R \\
\frac{tc}{r} \left( \ln \frac{tc/R}{1-r/R} - 1 \right) + \frac{1-r/R}{r/R}\ , \\
  \ \ \ \ \ \ \ \ \ {\rm if:}\ \ \  1{-}r/R{<}tc/R{<}1{+}r/R  \\
\frac{tc}{r} \ln \frac{1+r/R}{1-r/R} - 2 \ ,\ \ \ \ {\rm if:}\ \ \ \ tc/R{>}1{+}r/R
\end{array}
\right.
\ee

The surface of the discontinuity is characterized by 
the $T(r,t) = T_c$ contour line. If $T_c$ is the ignition temperature, then
here the DT ignition takes place on this contour line in the space-time.
The tangent of this line is if $tc>R+r$ :
\begin{multline}
 {\left( \frac{\partial r}{c\partial t} \right)}_{ T_c} = 
 {\left( \frac{\partial T}{c\partial t} \right)}_{ T_c} \Bigg/ 
 {\left( \frac{\partial T}{\partial r} \right)}_{ T_c} \\ 
 = { \ln \frac{1+r/R}{1-r/R} } \Bigg/ 
 \left\{ { { \left\lbrack \frac{2tc}{R-r}-\frac{tc}{r} \ln  
 \frac{1+r/R}{1-r/R} \right\rbrack} } \right\}
\end{multline}
So the point $(r_c, t_c)$ where the spacelike and timelike 
parts of the surface meet:
\begin{equation}
{\left( \frac{\partial r}{c\partial t} \right)}_{ T_c} \!\!= 
1 \rightsquigarrow t_c = 
{\left\{\frac{2c}{R-r_c}\Bigg[ \ln 
\frac{1+r_c/R}{1-r_c/R}\phantom{^I}\Bigg]^{-1} 
\!\!\!\! - \frac{c}{r_c} \right\} }^{-1}
\end{equation}

This line $t = t_c(r_c)$ separates the Space-like and Time-like branch 
of the discontinuity of $T(r,t) = T_c$.
\begin{figure}[!h]
\begin{center}
\resizebox{0.834\columnwidth}{!}
{\includegraphics{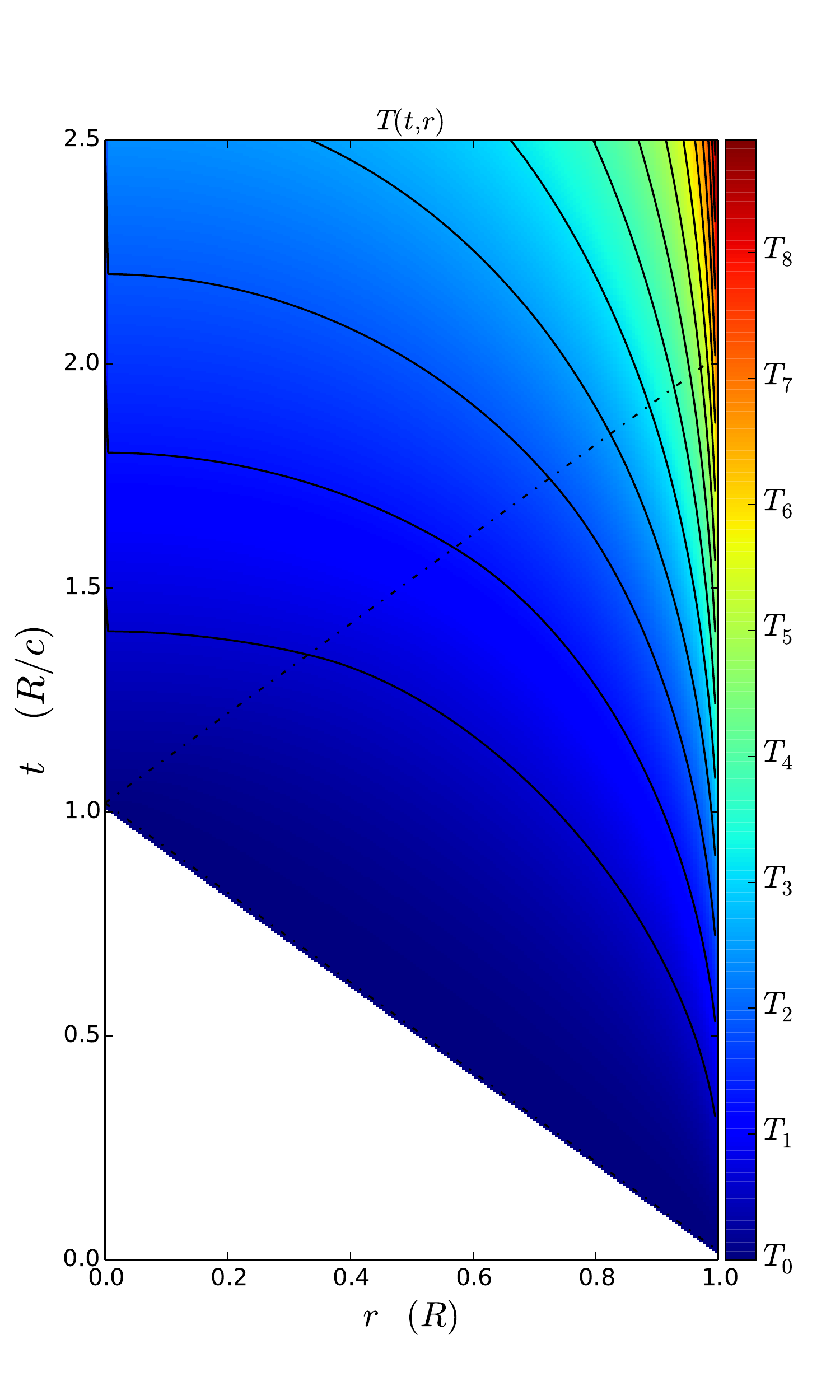}}
\caption{
(color online) 
The temperature distribution in function of distance and time. 
The dotted lines represent the light cone. The temperature is 
measured in units of  
$T_1 = H\cdot R= 272\,{\rm keV} $,
and $T_n = n \cdot T_1$.}
\label{F-3}
\end{center}
\end{figure}
The discontinuity initiates at $r=R$ and $t=0$ and it propagates 
first slowly inwards. Due to the radiative heat transfer 
the contour line of ignition, $T(r,t) = T_c$,
accelerates inwards and at $r_c = r_c (T_c)$ it develops smoothly 
from space-like into a time-like discontinuity.

The same type of gradual development from space-like into time-like 
detonation may occur in the last phase of ultra-relativistic heavy ion 
collisions. If we include radiative heat transfer in a scenario described 
in our ref., the transition from space-like 
to time-like deflagration will be gradual. This, however, 
requires more involved numerical calculations.

\section{Conclusions and discussions}

In this model estimate, we have neglected the compression of the 
target solid fuel ball, as well as the reflectivity of the target matter.   
The relatively small absorptivity made it possible that the radiation
could penetrate the whole target.  With the model parameters we used
the characteristic temperature was $T_1 = 272\,{\rm keV} $, which 
is largr than the usually assumed ignition temperature, while 
our target is not compressed so the higher temperatures may be
necessary to reach ignition according to the Lawson criterion.
If we can achieve ignition at somewhat lower temperature than $T_1$,
the ignition surface in the space time includes a substantial
time-like hyper-surface, where instabilities cannot develop, because
neighboring points are not causally connected. 

From looking at the temperature lines in function of distance 
and time (figure \ref{F-3}), we see that the detonation at a 
higher critical temperatures,
$T_c > T_3$  occurs
after the radiation reaches from the other side.  In this case, in the
outside domain of the target, the
inward propagation of the ignition front is slower than the speed of light 
(the dotted line on the figure indicates the speed of light reaching from the
other side of the pellet).  This makes it possible the development 
of instabilities
in this region.

The most optimal configuration is if we achieve ignition at $T_c = T_1$ or
slightly below. This leads to the fastest complete ignition of the target
with the least possibility that instabilities occur.

An alternative possibility to apply this model is to consider a
pre-compressed, more dense target, which is transparent and 
larger absorptivity. In this situation the ignition temperature is
somewhat smaller, but we still can optimize the pulse strength
and pulse length to achieve the fastest complete ignition of the 
target.

We can see if we neglect the importance of the speed of light, 
the theory would be far-fetched from reality. 
It is important to use the proper relativistic treatment
to optimize the fastest completer ignition, with the least possibility of 
instabilities, which reduce the efficiency of ignition.

\section*{Acknowledgements}

Enlightening discussions with Norbert Kroo are gratefully acknowledged.
This work is supported in part by the Institute of Advance Studies, 
K{\H o}szeg, Hungary and by the Collaboration Project between BKK and the
University of Bergen.



\begin{thebibliography}{99}


\bibitem{Lindl1998}
J. D. Lindl,  {\it Inertial Confinement Fusion} (Springer, 1998).

\bibitem{Lindl2004}
John D. Lindl, Peter Amendt, Richard L. Berger, S. Gail Glendinning, 
Siegfried H. Glenzer, Steven W. Haan, Robert L. Kauffman, Otto L. 
Landen1 and Laurence J. Suter,
The physics basis for ignition using indirect-drive targets on 
the National Ignition Facility,
Phys. Plasmas {\bf 11}, 339 (2004).

\bibitem{Haan2011}
S. W. Haan, J. D. Lindl, D. A. Callahan, D. S. Clark, J. D. Salmonson, 
B. A. Hammel, L. J. Atherton, R. C. Cook, M. J. Edwards, S. Glenzer, 
A. V. Hamza, S. P. Hatchett, M. C. Herrmann, D. E. Hinkel, D. D. Ho, 
H. Huang, O. S. Jones, J. Kline4, G. Kyrala, O. L. Landen, 
B. J. MacGowan, M. M. Marinak, D. D. Meyerhofer, J. L. Milovich, 
K. A. Moreno, E. I. Moses, D. H. Munro, A. Nikroo, R. E. Olson, 
K. Peterson, S. M. Pollaine, J. E. Ralph, H. F. Robey, B. K. Spears, 
P. T. Springer, L. J. Suter, C. A. Thomas, R. P. Town, R. Vesey, 
S. V. Weber, H. L. Wilkens, and D. C Wilson,
Point design targets, specifications, and requirements for 
the 2010 ignition campaign on the National Ignition Facility
Phys. Plasmas {\bf 18}, 051001 (2011).

\bibitem{RBOAH16}
R. Betti and O.A. Hurricane, Inertial-confinement fusion with lasers.
Nature Physics {\bf 12}, 435 (2016). 

\bibitem{Nora2015}
R. Nora, W. Theobald, R. Betti, F. J. Marshall, D.T. Michel, 
W. Seka, B. Yaakobi, M. Lafon, C. Stoeckl, J. Delettrez, A.A. Solodov, 
A. Casner, C. Reverdin, X. Ribeyre, A. Vallet, J. Peebles, F.N. Beg, 
and M.S. Wei,
Gigabar Spherical Shock Generation on the OMEGA Laser,
Phys. Rev. Lett. {\bf 114}, 045001(2015).

\bibitem{NIF11}
 D. S. Clark, M. M. Marinak, C. R. Weber, D. C. Eder, S. W. Haan, 
 B. A. Hammel, 
 D. E. Hinkel, O. S. Jones, J. L. Milovich, P. K. Patel, H. F. Robey, 
 J. D. Salmonson, 
 S. M. Sepke, and C. A. Thomas, 
Radiation hydrodynamics modeling of the highest compression 
inertial confinement fusion ignition experiment from the 
National Ignition Campaign,
Phys. Plasmas {\bf 22}, 022703 (2015).

\bibitem{RHL16}
V.H. Reis, R.J. Hanrahan, W.K. Levedahl, 
The big science of stockpile stewardship.
Physics Today {\bf 69}, 46 (2016).

\bibitem{HuCoGo14}
S.X. Hu, L.A. Collins, V.N. Goncharov, T.R. Boehly, R. Epstein, 
R.L. McCrory, and S. Skupsky,
First-priciples opacity table of warm dense deuterium for 
inertial-confinement-fusion applications.
Phys. Rev. E {\bf 90}, 033111 (2014).	

\bibitem{LCJ16}
L.C. Jarrott, M.S. Wei, C. McGuffey, A.A. Solodov, W. Theobald, 
B. Qiao, C. Stoeckl, R. Betti, H. Chen, J. Delettrez, T. D\"oppner, 
E.M. Giraldez, V.Y. Glebov, H. Habara, T. Iwawaki, M.H. Key, R.W. Luo, 
F.J. Marshall, H.S. McLean, C. Mileham, P.K. Patel, J.J. Santos, 
H. Sawada, R.B. Stephens, T. Yabuuchi, and F.N. Beg, 
Visualizing fast electron energy transport into 
laser-compressed high density fast-ignition targets.
Nature Physics {\bf 12}, 499 (2016).

\bibitem{C87}
L.P. Csernai,
 Detonation on a time-like front for relativistic systems,
Zh. Eksp. Teor. Fiz. {\bf 92}, 379-386 (1987).

\bibitem{CS14}
L.P. Csernai and D.D. Strottman,
Volume ignition via time-like detonation in pellet fusion
Laser and Particle Beams {\bf 33}, 279-282 (2015).		


\end{thebibliography}
\end{document}